\documentclass{svmult}
\usepackage{mathptmx}
\usepackage{tabularx}
\usepackage[dvips]{graphicx}
\graphicspath{{figures/}}
\DeclareGraphicsExtensions{.eps}
\usepackage{color}
\definecolor{grey}{rgb}{0.95, 0.95, 0.95}
\usepackage[final]{listings}
\usepackage{array}
\usepackage{textcomp}
\usepackage[T1]{fontenc}
\usepackage{tikz}
\usetikzlibrary{calc,arrows,patterns}
\tikzset{>=latex}
\usepackage{verbatim}
\usepackage{pgfplots, pgfplotstable}

\begin{document}

\title{Validation of hardware events for successful performance pattern identification in High Performance Computing}

\author{Thomas R\"ohl\thanks{Parts of this work were funded by the German Federal Ministry of Research and Education (BMBF) under Grant Number 01IH13009.}, Jan Eitzinger, Georg Hager, Gerhard Wellein\\thomas.roehl\,|\,jan.eitzinger\,|\,georg.hager\,|\,gerhard.wellein@fau.de}
\institute{Erlangen Regional Computing Center (RRZE)\\
University of Erlangen-Nuremberg\\
Erlangen, Germany}
\titlerunning{Validation of hardware events for performance pattern identification in HPC}
\authorrunning{Thomas R\"ohl, Jan Eitzinger, Georg Hager, Gerhard Wellein}

\maketitle

\abstract{Hardware performance monitoring (HPM) is a crucial
  ingredient of performance analysis tools. While there are
  interfaces like LIKWID, PAPI or the kernel interface perf\_event which provide HPM access with some additional features,
  many higher level tools combine event counts with results retrieved
  from other sources like function call traces to derive
  (semi-)automatic performance advice. However, although HPM is
  available for x86 systems since the early 90s, only a small subset
  of the HPM features is used in practice. Performance patterns
  provide a more comprehensive approach, enabling the identification
  of various performance-limiting effects. Patterns address issues
  like bandwidth saturation, load imbalance, non-local data access in
  ccNUMA systems, or false sharing of cache lines. This work defines
  HPM event sets that are best suited to identify a selection of
  performance patterns on the Intel Haswell processor. We validate
  the chosen event sets for accuracy in order to arrive at a reliable
  pattern detection mechanism and point out shortcomings that cannot
  be easily circumvented due to bugs or limitations in the hardware.}

\section{Introduction and related work}
Hardware performance monitoring (HPM) was introduced for the x86 architecture with the Intel Pentium in 1993 \cite{ryan1993inside}.
Since that time, HPM gained more and more attention in the computer science community and consequently a lot of HPM related tools were developed.
Some provide basic access to the HPM registers with some additional features like LIKWID \cite{likwid}, PAPI \cite{mucci1999papi} or the kernel interface perf\_event~\cite{perfevent}.
Furthermore, some higher level analysis tools gather additional information by combining the HPM counts with application level traces.
Popular representatives of that analysis method are HPCToolkit~\cite{adhianto2010hpctoolkit}, PerfSuite~\cite{kufrin2005perfsuite}, Open|Speedshop~\cite{schulz2008open} or Scalasca~\cite{geimer2010scalasca}.
The intention of these tools is to advise the application developer with educated optimization hints.
To this end, the tool developers use performance metrics that represent a possible performance limitation, such as saturated memory bandwidth or instructions paths.
The hardware metrics may be combined with information on the application level, e.g. scaling characteristics, dependence of performance on the problem size, or static code analysis, to arrive at a \emph{signature}.
A performance signature then points towards one or more \emph{performance patterns}, as described in \cite{patterns} and refined in \cite{patterns-poster}.
The purpose of the patterns concept is to facilitate the identification of performance-limiting bottlenecks.

C. Guillen uses in \cite{guillen} the term \emph{execution properties} instead of performance pattern.
She defines execution properties as a set of values gathered by monitoring and related thresholds.
The properties are arranged in decision trees for compute- and memory-bound applications as well as trees related to I/O and other resources.
This enables either a guided selection of the analysis steps to further identify performance limitation or automatic tool-based analysis.
Based on the path in the decision tree, suggestions are given for what to look for in the application code.
A combination of the structured performance engineering process in \cite{patterns-poster} with the decision trees in \cite{guillen} defines a good basis for (partially) automated performance analysis tools.

One main problem with HPM is that none of the main vendors for x86 processors guarantees event counts to be accurate or deterministic.
Although many HPM interfaces exist, only little research has been done on validating the hardware performance events. However, users tend to trust the returned HPM counts and use them for decisions about code optimization.
One should be aware that HPM measurements are only guideposts until the HPM events are known to have guaranteed behavior. Moreover, analytic performance models can only be validated if this is the case.

The most extensive event validation analysis was done by Weaver et al.\ \cite{Weaver2013} using a self-written assembly validation code.
They test determinism and overcounting for the following events: retired instructions, retired branches, retired loads and stores as well as retired floating-point operations including scalar, packed, and vectorized instructions.
For validating the measurements the dynamic binary instrumentation tool \emph{Pin} \cite{Luk:2005:PBC:1064978.1065034} was used.
The main target of that work was not to identify the right events needed
to construct accurate performance metrics but to find the sources of non-determinism and over/undercounting.
It gives hints on how to reduce over- or undercounting and identify deterministic events for a set of architectures.

D. Zaparanuks et al.\ \cite{4919635} determined the error of retired instructions and CPU cycle counts with two microbenchmarks.
Since the work was released before the perf\_event interface \cite{perfevent} was available for PAPI, they tested the deprecated interfaces perfmon2 \cite{eranian2006perfmon2} and perfctr \cite{pettersson2003perfctr} as the basis for PAPI.
They use an ``empty'' microbenchmark to define a default error using different counter access methods.
For subsequent measurements they use a simple loop kernel with configurable iterations, define a model for the code and compare the measurement results to the model.
Moreover, they test whether the errors change for increasing measurement duration and for a varying number of programmed counter registers.
Finally, they give suggestions which back-end should be used with which counter access pattern to get the most accurate results.

In the remainder of this section we recommend HPM event sets and related derived metrics that represent the signature of prototypical examples picked out of the performance patterns defined in \cite{patterns-poster}.
In the following sections the accuracy of the chosen HPM events and their derived metrics is validated.
Our work can be seen as a recommendation for tool developers which event sets match the selected performance patterns in the best way and how reliable they are.

\section{Identification of signatures for performance patterns}
Performance patterns help to identify possible performance problems in an application.
The measurement of HPM events is one part of the pattern's signature.
There are patterns that can be identified by HPM measurements alone, but commonly more information is required, e.g., 
scaling behavior or behavior with different data set sizes.
Of course, some knowledge about the micro-architecture is also required to select the proper event sets for HPM as well as to determine the capabilities of the system.
For x86 systems, HPM is not part of the instruction set architecture (ISA), thus besides a few events spanning multiple micro-architectures, each processor generation defines its own list of HPM events.
Here we choose the Intel Haswell EP platform (E5-2695 v3) for HPM event selection and verification. The general approach can certainly be applied to other architectures.

In order to decide which measurement results are good or bad, the characteristics of the system must be known.
C. Guillen established thresholds in \cite{guillen} with four different approaches: hardware characteristics, expert knowledge about hardware behavior and performance optimization, benchmarks and statistics.
With decision trees but without source code knowledge it is possible to give some loose hints how to further tune the code.
With additional information about the software code and run time behavior, the list of hints could be further reduced.

The present work is intended to be a referral for which HPM events provide the best information to specify the signatures of the selected performance patterns.
The patterns target different behaviors of an application and/or the hardware and therefore are classified in three groups: \emph{bottlenecks}, \emph{hazards} and \emph{work-related} patterns.
The whole list of performance patterns with corresponding event sets for the Intel Haswell EP micro-architecture can be found at \cite{patterns-wiki}.
For brevity we restrict ourselves to three patterns: \emph{bandwidth saturation}, \emph{load imbalance} and \emph{false sharing of cache lines}.
For each pattern, we list possible signatures and shortcomings concerning the coverage of a pattern by the event set.
The analysis method is comparable to the one of D. Zaparanuks et al. \cite{4919635} but uses a set of seven assembly benchmarks and synthetic higher level benchmark codes that represent often used algorithms in scientific applications.
But instead of comparing the raw results, we use derived metrics, combining multiple counter values, for comparison as these metric results are commonly more interesting for tool users.

\subsection*{Bandwidth saturation}
A very common bottleneck is bandwidth saturation in the memory hierarchy, notably at the memory interface but also in the L3 cache on earlier Intel designs. Proper identification of this pattern requires an accurate measurement of the data volume, i.e., the number of transferred cache lines between memory hierarchy levels.
From data volume and run time one can compute transfer bandwidths, which can then be compared with measured or theoretical upper limits.

Starting with the Intel Nehalem architecture, Intel separates a CPU socket in two components, the core and the uncore.
The core embodies the CPU cores and the L1 and L2 caches.
The uncore covers the L3 cache as well as all attached components like memory controllers or the Intel QPI socket interconnect.
The transferred data volume to/from memory can be monitored at two distinct uncore components.
A CPU socket in an Intel Haswell EP machine has at most two memory controllers (iMC) in the uncore, each providing up to four memory channels.
The other component is the Home Agent (HA) which is responsible for the protocol side of memory interactions.

Starting with the Intel Sandy Bridge micro-architecture, the L3 cache is segmented, with one segment per core. Still one core can make use of all segments. 
The data transfer volume between the L2 and L3 caches can be monitored in two different ways: One may either count the cache lines that are requested and written back by the L2 cache, or the lookups for data reads and victimized cache lines that enter the L3 cache segments.
It is recommended to use the L2-related HPM events because the L3 cache is triggered by many components besides the L2 caches.
Moreover, the Intel Haswell EP architecture has up to $18$ L3 cache segments which all need to be configured separately.
Bandwidth bottlenecks between L1 and L2 cache or L1 and registers are seldom and thus ignored in this pattern.

\subsection*{Load imbalance}
The main characterization of this pattern is that different threads have to process different working sets between synchronization points.
For data-centric workloads the data volume transferred between the L1 and L2 caches for each thread may be an indicator: since the working sets have different sizes, it is likely that smaller working sets also require less data.
However, the assumption that working set size is related to transferred cache lines is not expressive enough to fully identify the pattern, since
the amount of required data could be the same for each thread while the amount of in-core instructions differs.
Retired instructions, on the other hand, are just as unreliable as data transfers because parallelization overhead often comprises spin-waiting loops that cause abundant instructions without doing ``work.''
Therefore, for better classification, it is desirable to count ``useful'' instructions that perform the actual work the application has to do.
None of the two x86 vendors provides features to filter the instruction stream and count only specific instructions in a sufficiently flexible way.
Moreover, the offered hardware events are not sufficient to overcome this shortcoming by covering most ``useful'' instructions like scalar/packed floating-point operations, SSE driven calculations or string related operations.
Nevertheless, filtering on some instruction groups works for Intel Haswell systems, such as long-latency instructions (div, sqrt,...) or AVX instructions.
Consequently, it is recommended to measure the work instructions if possible but also the data transfers can give a first insight.

\subsection*{False cache line sharing}
False cache line sharing occurs when multiple cores access the same cache line while at least one is writing to it.
The performance pattern thus has to identify bouncing cache lines between multiple caches.
There are codes that require true cache line sharing, like producer/consumer codes, but we are referring to common HPC codes where cache line sharing should be as minimal as possible.
In general, the detection of false cache line sharing is very hard when restricting the analysis space only to hardware performance measurements.
The Intel Haswell micro-architecture offers two options for counting cache line transfers between private caches:
There are L3 cache related $\mu$OPs events for intra- and inter-socket transfers, but the HPM event for intra-socket movement may undercount with SMT enabled by as much as $40\%$ according to erratum HSW150 in \cite{haswell-specs}.
The alternative is the \emph{off\-core response unit}.
By setting the corresponding filter bits, the L3 hits with hitm snoops (hit a modified cache line) to other caches on the socket and the L3 misses with hitm snoops to remote sockets can be counted.
The specification update \cite{haswell-specs} also lists an erratum for the offcore response unit (HSW149) but the required filter options for shared cache lines are not mentioned in it.
There are no HPM events to count the transfers of shared cache lines at the L2 cache.
In order to clearly identify whether a code triggers true or false cache line sharing, further information like source code analysis is required.

\section{Useful event sets}

\begin{table}[bt]
\caption{\label{pattern_table}Desired events, available events and comments for three performance patterns on the Intel Haswell EP micro-architecture. A complete list can be found at \cite{patterns-wiki}}
\label{pattern-events}
\begin{tabular}{|p{\dimexpr 0.3\linewidth-2\tabcolsep}|p{\dimexpr 0.3\linewidth-2\tabcolsep}|p{\dimexpr 0.4\linewidth-2\tabcolsep}|}
\hline 
Pattern & Desired events & Available events  \\ 
\hline 
Bandwidth saturation & Data volume transferred to/from memory from/to the last level cache; data volume transferred between L2 and L3 cache & iMC:{\tt UNC\_M\_CAS\_COUNT.RD}, iMC:{\tt UNC\_M\_CAS\_COUNT.WR}, HA:{\tt UNC\_H\_IMC\_READS.NORMAL}, HA:{\tt UNC\_H\_BYPASS\_IMC.TAKEN}, HA:{\tt UNC\_H\_IMC\_WRITES.ALL}, {\tt L2\_LINES\_IN.ALL}, {\tt L2\_TRANS.L2\_WB}, CBOX:{\tt LLC\_LOOKUP.DATA\_READ}, CBOX:{\tt LLC\_VICTIMS.M\_STATE}  \\ 
\hline 
Load imbalance & Data volume transferred at all cache levels; number of ``useful'' instructions & {\tt L1D.REPLACEMENT}, {\tt L2\_TRANS.L1D\_WB}, {\tt L2\_LINES\_IN.ALL}, {\tt L2\_TRANS.L2\_WB}, {\tt AVX\_INSTS.CALC}, {\tt ARITH.DIVIDER\_UOPS} \\
\hline 
False sharing of cache lines & All transfers of shared cache lines for the L2 and L3 cache; all transfers of shared cache lines between the last level caches of different CPU sockets & {\tt MEM\_LOAD\_UOPS\_L3\_ HIT\_RETIRED.XSNP\_HITM}, {\tt MEM\_LOAD\_UOPS\_L3\_ MISS\_RETIRED.REMOTE\_HITM}, {\tt OFFCORE\_RESPONSE: LLC\_HIT:HITM\_OTHER\_CORE}, {\tt OFFCORE\_RESPONSE: LLC\_MISS:REMOTE\_HITM} \\
\hline 
\end{tabular} 
\end{table}

Table~\ref{pattern_table} defines a range of HPM event sets that are best suitable for the described performance patterns regarding the HPM capabilities of the Intel Haswell EP platform.
The assignment of HPM events for the pattern signatures is based on the Intel documentation (\cite{sdm}, \cite{haswell-uncore}).
Some events are not mentioned in the default documentation; they are taken from Intel's performance monitoring database \cite{perfmondb}.
Although the events were selected with due care, there is no official guarantee for the accuracy of the counts by the manufacturer.
The sheer amount of performance monitoring related errata for the Intel Haswell EP architecture \cite{haswell-specs} reduces the confidence even further.
But this encourages us even more to validate the chosen event sets in order to provide tool developers and users a reliable basis for their performance analysis.


\section{Validation of performance patterns}
Many performance analysis tools use the HPM features of the system as their main source of information about a running program. They assume event counts to be correct, and some even generate automated advice for the developer.
Previous research in the field of HPM validation focuses on singular events like retired instructions but does not verify the results for other metrics that are essential for identifying performance patterns.
Proper verification requires the creation of benchmark code that has well-defined and thoroughly understood performance features and, thus, predictable event counts.
Since optimizing compilers can mutilate the high level code, the feasible solutions are either to write assembly benchmarks or to perform code analysis of the assembly code created by the compiler.

The LIKWID tool suite \cite{likwid} includes the {\tt likwid-bench} microbenchmarking framework, which provides a set of assembly language kernels. They cover a variety of streaming access schemes.
In addition the user can extend the framework by writing new assembly code loop bodies.
\texttt{likwid-bench} takes care of loop counting, thread parallelism, thread placement, ccNUMA page placement and performance (and bandwidth) measurement.
It does not, however, perform hardware event counting.
For the HPM measurements we thus use {\tt likwid-perfctr}, which is also a part of the LIKWID suite.
It uses a simple command line interface but provides a comprehensive set of features for the users.
{\tt Likwid-perfctr} supports almost all interesting core and uncore events for the supported CPU types.
In order to relieve the user from having to deal with raw event counts, it supports \emph{performance groups}, which combine often used event sets and corresponding formulas for computing derived metrics (e.g., bandwidths or FLOP rates).
Moreover, {\tt likwid-perfctr} provides a \emph{Marker API} to instrument the source code and restrict measurements to certain code regions.
{\tt Likwid-bench} already includes the calls to the Marker API in order to measure only the compute kernel.
We have to manually correct some of the results of {\tt likwid-bench} to represent the obvious and hidden data traffic (mostly write-allocate transfers) that may be measured with {\tt likwid-perfctr}. 

\pgfplotstableread{
Time Value Min Max
0 0.475 0.29 0.73
1 0.215 0.14 0.32
2 0.37 0.21 0.53
3 1.26 0.9 1.67
4 1.86 1.52 2.17
5 2.73 2.23 3.13
6 0.05 0.04 0.08
7 -0.2  -0.17 -0.27
8 -1.1075 -0.96 -1.19
9 -0.115 -0.01 -0.18
10 0.035 0.01 0.09
11 -6.3475 -6.22 -6.44
12 0.21 0.2 0.23
13 0.16  0.15 0.18
14 0.165 0.15 0.21
15 0.1675 0.15 0.21
16 0.16 0.15 0.19
17 0.16 0.15 0.2
18 0.15 0.14 0.21
19 0.1225 0.11 0.17
20 0.12 0.11 0.16
21 0.12 0.09 0.15
22 0.125 0.11 0.17
23 0.1125 0.09 0.13
}\datatableall

\pgfplotsset{
    min_err/.style={
        mark=empty,
        error bars/.cd,
            y dir=minus,
            y explicit,
        /pgfplots/table/.cd,
            x=Time,
            y=Value,
            y error expr=\thisrow{Value}-\thisrow{Min}
    },
    max_err/.style={
        mark=empty,
        error bars/.cd,
            y dir=plus,
            y explicit,
        /pgfplots/table/.cd,
            x=Time,
            y=Value,
            y error expr=\thisrow{Max}-\thisrow{Value}
    }
}

\begin{figure}[tb]
\begin{tikzpicture}
\begin{axis}[width=\textwidth, only marks,ylabel={Error in \%}, xticklabels={L2\_load,L2\_store,L2\_copy,L2\_stream,L2\_daxpy,L2\_triad, L3\_load,L3\_store,L3\_copy,L3\_stream,L3\_daxpy,L3\_triad,MEM\_load,MEM\_store,MEM\_copy,MEM\_stream,MEM\_daxpy,MEM\_triad,HA\_load,HA\_store,HA\_copy,HA\_stream,HA\_daxpy,HA\_triad}, xtick={0,...,23},
x tick label style={rotate=90,anchor=east}]
\addplot[min_err] table {\datatableall};
\addplot[max_err] table {\datatableall};
\addplot[mark=*] table[x index={0}, y index={1}] {\datatableall};
\end{axis}
\end{tikzpicture}
\caption{Verification tests for cache and memory traffic using a set of micro benchmarking kernels written in assembly. We show the average, minimum and maximum error in the delivered HPM counts for a collection of streaming kernels with data in L2, L3 and in memory.}
\label{band_sat}
\end{figure}
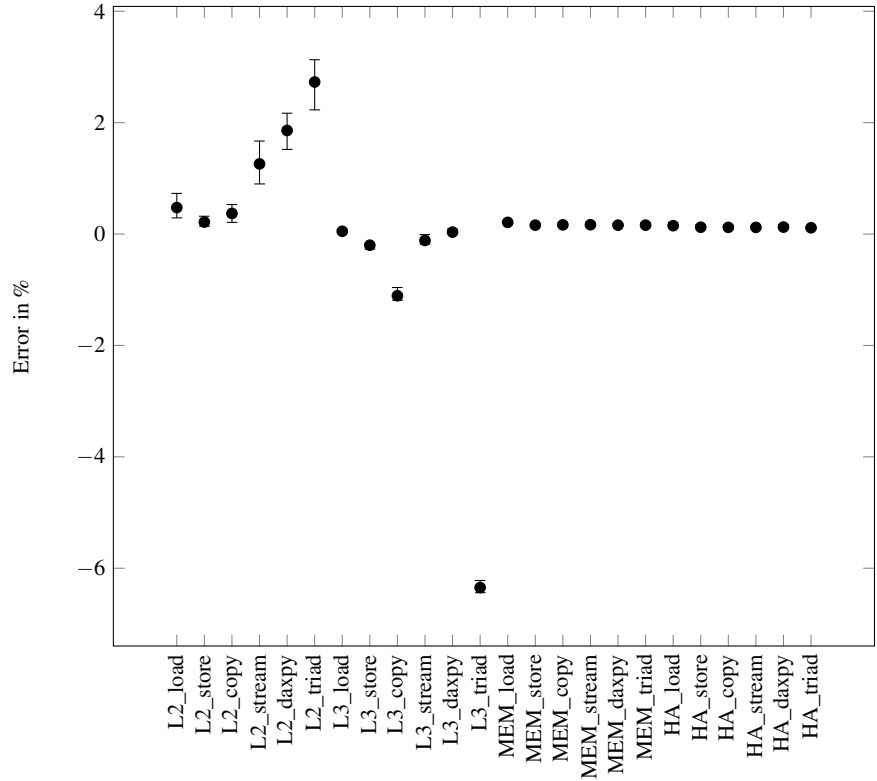

The first performance pattern for the analysis is the bandwidth saturation pattern.
For this purpose, {\tt likwid-perfctr} already provides three performance groups called L2, L3 and MEM \cite{likwid}.
A separate performance group was created to measure the traffic traversing the HA.
Based on the raw counts, the groups define derived metrics for data volume and bandwidth.
For simplicity we use the derived metric of total bandwidth for comparison as it both includes the data volume in both directions and the run time.
In Fig.~\ref{band_sat} the average, minimal and maximal errors of $100$ runs with respect to the exact bandwidth results are presented for seven streaming kernels and data in L2 cache, L3 cache and memory.
The locality of the data in the caching hierarchy is ensured by streaming accesses to the vectors fitting only in the relevant hierarchy level.
The first two kernels (\emph{load} and \emph{store}) perform pure loading and storing of data to/from the CPU core to the selected cache level or the memory.
A combination of both is applied in the \emph{copy} test.
The last three tests are related to scientific computing and well understood.
They range from the linear combination of two vectors called \emph{ddot} calculating $A[i] = B[i] \cdot c + A[i]$, a \emph{stream} triad with formula $A[i] = B[i] \cdot c + C[i]$ to a vector \emph{triad} computing $A[i] = B[i] \cdot C[i] + D[i]$.

The next pattern we look at is load imbalance. Since load imbalance requires a notion of ``useful work'' we have to find a way to measure floating-point operations.
Unfortunately, the Intel Haswell architecture lacks HPM events to fully represent FLOP/s.
For the Intel Haswell architecture, Intel has documented a HPM event {\tt AVX\_INSTS.ALL} (Event 0xC6, Umask 0x07) which captures all AVX instructions including data movement and calculations \cite{perfmondb}.
With the help of {\tt likwid-bench} we could further refine the event to count loads (Umask 0x01), stores (Umask 0x02) and calculations (Umask 0x04) separately.
Consequently, the FLOP/s performed with AVX operations can be counted.
All performance patterns that require the filtering of the instruction stream for specific instructions can use the event {\tt AVX\_INSTS.CALC} for floating-point operations using the AVX vectorization extension.
Due to its importance, the event is verified using the {\tt likwid-bench} utility with assembly benchmarks that are based on AVX instructions only.
Note that the use of these specific Umasks is an undocumented feature and may change with processor generations or even mask revisions.
Moreover, we have found no way to count SSE or scalar floating-point instructions.

\pgfplotstableread{
Time Value Min Max
0 0.045 0.01 0.13
1 0.04 0.01 0.11
2 0.03 0.01 0.08
3 0.065 0.01 0.16
}\datatableflops

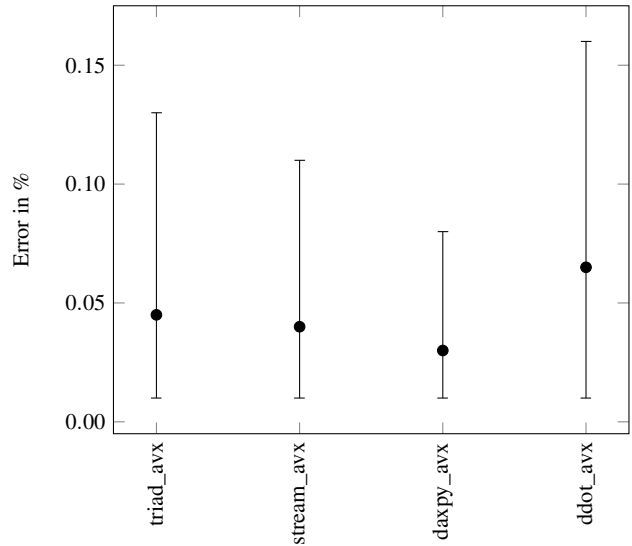
\begin{figure}[tb]
\begin{tikzpicture}
\begin{axis}[only marks,ylabel={Error in \%}, xticklabels={triad\_avx, stream\_avx, daxpy\_avx, ddot\_avx}, xtick={0,...,3},
x tick label style={rotate=90,anchor=east}, y tick label style={/pgf/number format/.cd,fixed,fixed zerofill,precision=2,/tikz/.cd}]
\addplot[min_err] table {\datatableflops};
\addplot[max_err] table {\datatableflops};
\addplot[mark=*] table[x index={0}, y index={1}] {\datatableflops};
\end{axis}
\end{tikzpicture}
\caption{Verification tests for the AVX floating point event using a set of microbenchmarking kernels with pure AVX code.}
\label{flops_avx}
\end{figure}
Fig.~\ref{flops_avx} shows the minimum, maximum and average error for measuring AVX FLOP/s.
The average error for all tests is below $0.07\%$. As the maximal error is $0.16\%$ the event can be seen as sufficiently accurate for pure AVX code.
Using the counter with non-AVX codes always returns 0.

Coming back to performance patterns, we now verify the load imbalance pattern using an upper triangular matrix vector multiplication code running with two threads.
Since the accuracy of the cache and memory traffic related HPM events have been verified already, we use the only available floating-point operation related event {\tt AVX\_INSTS.CALC}.
There is one shortcoming worth noting:
If the code contains half-wide loads, the HPM event shows overcounting.
The compiler frequently uses half-wide loads to reduce the probability of ``split loads,'' i.e., AVX loads that cross a cache line boundary if 32-byte alignment cannot be guaranteed.
Experiments have shown that the event {\tt AVX\_INSTS.CALC} includes the {\tt vinsertf128} instruction as a calculation operation.
In order to get reliable results, split AVX loads should be avoided.
This is not a problem with \texttt{likwid-bench} as no compiler is involved and the generated assembly code is under full control.
The upper triangular matrix is split so that each of the two threads operates on half of the matrix.
The matrix has a size of $8192 \times 8192$ and the multiplication is performed $1000$ times.
The first thread processes the top rows with totally $25167872$ elements, while the second one works on the remaining $8390656$ elements.
This distribution results in a work load imbalance for the threads of $3:1$.
\begin{table}[tb]
\caption{Verification of the load imbalance pattern using an upper triangular matrix vector multiplication code}
\begin{tabular}{|p{\dimexpr 0.34\linewidth-2\tabcolsep}|>{\centering\arraybackslash}m{\dimexpr 0.165\linewidth-2\tabcolsep}|>{\centering\arraybackslash}m{\dimexpr 0.165\linewidth-2\tabcolsep}|>{\centering\arraybackslash}m{\dimexpr 0.165\linewidth-2\tabcolsep}|>{\centering\arraybackslash}m{\dimexpr 0.165\linewidth-2\tabcolsep}|}
\hline 
Event/Metric & Thread 0 & Thread 1 & Ratio & Error\\ 
\hline 
Process elements & $25167872$ & $8390656$ & $3:1$ &  \\ 
\hline
AVX floating point ops & $1.26\mbox{\textsc{e}}10$ & $4.21\mbox{\textsc{e}}09$  & $2.991:1$ & $0.29\%$\\ 
\hline
L2 data volume [GByte] & $406.28$ & $115.34$  & $3.52:1$ & $17.42\%$ \\ 
\hline
L3 data volume [GByte] & $203.06$ & $69.74$  & $2.912:1$ & $2.94\%$ \\ 
\hline 
Memory data volume [GByte] & $112.97$ & $ 37.33$  & $3.026:1$ & $0.88\%$ \\ 
\hline 
\end{tabular}
\label{tab-load-imbalance}
\end{table}

Table~\ref{tab-load-imbalance} lists the verification data for the code.
The AVX calculation instruction count fits to a high degree the work load ratio of $3:1$.
The L2 data volume has the highest error, mainly caused by repeatedly fetching the input and output vector not included in the work load balance model.
This behavior also occurs for the L3 and memory data volume but to a lesser extent as the cache lines of the input vector commonly stay in the caches.
In order to get the memory data volume per core, the offcore response unit was used.

The false sharing of cache lines pattern is difficult to verify as it is not easy to write code that shows a predictable number of inter-core cache line transfers.
A minimal amount of shared cache lines exist in almost every code thus HPM results unequal zero cannot be accepted as clear signature.
To measure the behavior, a producer and consumer code was written, thus we verify the amount of falsely shared cache lines by using a true sharing cache line code.
The producer writes to a consecutive range of memory that is read afterwards by the consumer.
In the next iteration the producer uses the subsequent range of memory to avoid invalidation traffic.
The memory range is aligned so that a fixed amount of cache lines is used in every step.
The producer and consumer perform $100$ iterations in each of the $100$ runs.
For synchronizing the two threads, a simple busy-waiting loop spins on a shared variable with long enough sleep times to avoid high access traffic for the synchronization variable.
When using pthread conditions and a mutex lock instead, the measured values are completely unstable.

\pgfplotstableread{
Lines Model MeasureLocal MeasureRemote
2 200 328.4 404.17
4 401 568.33 607.66
8 800 897 985.86
16 1601 1714.2 1777.14
32 3200 2893.4 2595.57
64 6400 5570.8 3512.29
128 12800 7350.66 6124.86
256 25600 8995.33 11471.14
512 51200 18224.29 22608.29
1024 102400 55124.5 45814.29
}\falseshare

\begin{table}[tb]
\caption{Verification tests for false sharing of cache lines using a producer/consumer code. The producer and consumer thread are located on the same CPU socket.}
\begin{tabular}{|>{\centering\arraybackslash}m{\dimexpr 0.19\linewidth-2\tabcolsep}|>{\centering\arraybackslash}m{\dimexpr 0.13\linewidth-2\tabcolsep}|>{\centering\arraybackslash}m{\dimexpr 0.2\linewidth-2\tabcolsep}|>{\centering\arraybackslash}m{\dimexpr 0.12\linewidth-2\tabcolsep}|>{\centering\arraybackslash}m{\dimexpr 0.2\linewidth-2\tabcolsep}|>{\centering\arraybackslash}m{\dimexpr 0.12\linewidth-2\tabcolsep}|}
\hline 
Amount of shared cache lines per step & Transferred cache lines according to model & Avg. amount of intra-socket transferred shared cache lines & Error [$\%$] & Avg. amount of inter-socket transferred shared cache lines & Error [$\%$]\\ 
\hline 
$2$ & $200$ & $328.4$ & $64.2$ & $404.2$ & $102.1$ \\ 
\hline
$4$ & $400$ & $568.3$ & $42.1$ & $607.7$ & $51.9$\\ 
\hline
$8$ & $800$ & $897.0$ & $12.1$ & $985.86$ & $23.2$\\
\hline
$16$ & $1600$ & $1714.2$ & $7.1$ & $1777.1$ & $11.1$ \\ 
\hline
$32$ & $3200$ & $2893.4$ & $-9.6$ & $2595.6$ & $-18.9$ \\ 
\hline 
$64$ & $6400$ & $5570.8$ & $-13.0$ & $3512.3$ & $-45.1$ \\
\hline 
$128$ & $12800$ & $7350.7$ & $-42.6$ & $6124.9$ & $-52.2$ \\
\hline 
$256$ & $25600$ & $8995.3$ & $-64.9$ & $11471.1$ & $-55.2$ \\
\hline 
$512$ & $51200$ & $18224.3$ & $-64.4$ & $22608.3$ & $-55.8$ \\
\hline 
$1024$ & $102400$ & $55124.5$ & $-46.2$ & $45814.3$ & $-55.3$ \\
\hline 
\end{tabular}
\label{false_share}
\end{table}


Table~\ref{false_share} shows the measurements for HPM events fitting best to the traffic caused by false sharing of cache lines.
The table lists the amount of cache lines that are written by the producer thread.
Since the consumer reads all these lines, the amount of transferred cache lines should be in the same range.
The measurements using the events in Tab.~\ref{pattern-events} show a big discrepancy between the counts in the model and the measured transfers.
For small counts of transferred cache lines, the results are likely to be distorted by the shared synchronization variable, but the accuracy should improve with increasing transfer sizes.
Since the erratum HSW150 in \cite{haswell-specs} states an undercounting by as much as $40\%$, the intra-socket measurements could be too low.
But even when scaling up the measurements the HPM event for intra-socket cache line sharing is not accurate.

For the inter-socket false sharing, the threads are distributed over the two CPU sockets in the system.
The results in Tab.~\ref{false_share} show similar behavior as in the intra-socket case.
The HPM events for cache line sharing provide a qualitative classification for the performance pattern's signature but no quantitative one.
The problem is mainly to define a threshold for the false-sharing rate of the system and application.
Further research is required to create suitable signature for this performance pattern.

\section{Conclusion}

The performance patterns defined in \cite{patterns-poster} provide a comprehensive collection for analyzing possible performance degradation on
the node level. They address possible hardware bottlenecks as well as typical inefficiencies in parallel programming.
We have listed suitable event sets to identify the bandwidth saturation, load imbalance, and false sharing patterns with HPM on the Intel Haswell architecture.
Unfortunately the hardware does not provide all required events, such as, e.g., scalar/packed floating-point operations, or they are not accurate enough like, e.g., the sharing of cache lines at the L3 level.
Moreover, a more fine-grained and correct filtering of instructions would be helpful for pattern-based performance analysis.

Using a selection of streaming loop kernels we found the error for the bandwidth-related events to be small on average ($-1\%\ldots {+2\%}$), with a maximum undercounting of about $-6\%$ for the L3 traffic.
The load imbalance pattern was verified using an upper triangular matrix vector multiplication.
Although the error for the L1 to L2 cache traffic is above $15\%$, the results reflect the correct load imbalance of roughly $3:1$, indicating the usefulness of the metrics.
Moreover, we have managed to identify filtered events that can accurately count AVX floating-point operations under some conditions.
FLOP/s and traffic data are complementary information for identifying load imbalance.
The verification of the HPM signature for the false sharing pattern failed due to large deviations from the expected event counts for the two events used.
More research is needed here to arrive at a useful procedure, especially for
distinguishing unwanted false cache line sharing from traffic caused by
intended updates. 

The remaining patterns defined in \cite{patterns-poster} need to be verified as well to provide a well-defined HPM analysis method for performance patterns ready to be included in performance analysis tools.
We provide continuously updated information about suitable events for pattern identification in the Wiki on the LIKWID website\footnote{https://github.com/RRZE-HPC/likwid}.

\bibliography{9ptw}
\bibliographystyle{acm}
\end{document}